# Fernando Sanford and the "Kirlian effect"

Amelia Carolina Sparavigna
Dipartimento di Fisica
Politecnico di Torino, Torino, Italy

In 1894 Fernando Sanford discussed in a paper published by the Physical Review an electric photography technique, which he developed starting from 1891. The images that he published in the paper are clearly showing some fringes about the objects he photographed. Several years after, the fringes appearing in the electric photography have been recognized as the "Kirlian effect". Unlike Kirlian, Sanford did his best to reduce these fringes, improving the device but losing the "Sanford effect".

Fernando Sanford was professor at the Stanford University. He was born on a farm near Franklin Grove, Illinois, February 12, 1856 [1]. After attaining his degree of Bachelor of Science, in 1879 he became Professor of Physical Science and then Superintendent of Schools in Illinois. To improve his knowledge of science, he went to Germany and studied for two years with one of the outstanding physicists of the time, Hermann Von Helmholtz. Though Von Helmholtz pressed him to stay in Berlin, Sanford came back in the United States, becoming one of the members of the group of scientists who came to Stanford to create its pioneer faculty in 1891.

Precisely from this year, Sanford started to study and test the electric photography. This was a photographic technique, experimented in that period such as other photographic "lightless techniques", by several researches. Reported on magazines, these researches stimulated also the popular culture and the literature [2].

Sanford describes the technique in a paper published on Physical Review [3], after becoming aware of the announcement of a new device, the "Inductoscript", invented by F.J. Smith. Sanford tells that in the earlier experiments, he developed a negative image upon a piece of bromide paper, that he placed between two plane metallic electrodes, connected with the poles of two chromic acid cells. The negatives showed only the places of contact of the sensitized side of the paper and the corresponding electrode. After reading a paper by W.B. Crofts, in the Philosophical Magazine of August, 1892, Sanford prepares an experiment laying a coin upon the sensitized side of a glass plate, and connecting it with the terminal of a small induction coil, capable of giving a spark, while a tin foil under the opposite side of the photographic plate was connected with the other terminal of the coil. Sanford is telling that this is the same method used in the "Inductoscript".

To demonstrate the effect of an electric photography, Sanford is accompanying the article with a photograph. Moreover, he tells that he did several negatives, and that, with only one exception, all show fringes around the photographed objects, fringes "due to the escape of the charge from the edge of the coin, which accounts for the formation of the dark ring observed around the breath figures made upon glass" (see Fig.1, left). An improvement of the set-up, obtained by using a dielectric insulation of the object, gives as a result the image on the right of Fig.1. In fact, Sanford wanted to remove the fringes from the image.

The photographic plates obtained by Sanford and shown in his article, are quite similar to those obtained by means of the Kirlian photography (see Fig.2). This is an electric photography too,

"named after Semyon Kirlian, who in 1939 accidentally discovered that if an object on a photographic plate is connected to a source of voltage an image is produced on the photographic plate" [3,4]. According to these references, Kirlian independently rediscovered the electric photography, the underlying physics of which was explored as early as 1777 by Georg Christoph Lichtenberg and then by several researchers, including Nikola Tesla.

What was enhancing the interest on electric photography was the Kirlian's assertion that the images he was obtaining might be due to the "aura" of a living being. Of course, in addition to living material, inanimate objects, such as coins, produce images on the film in a Kirlian photograph setup (see Figure 2). Reference [3] is reporting several studies on the Kirlian photography. Let me just cite the researches by K. Korotkov in St. Petersburg. He uses a Gas Discharge Visualization technique, parent of the Kirlian effect, where electrodes create a pulsed electric field perturbation that has a glowing effect. According to the researcher, this effect could be useful in some cases to detect the stress of biological structures.

A Google search for "Kirlian photography" produces about 278,000 results. Searching for "Kirlian aura" gives about 215,000 results. It means that the fringe effect in an electric photography is able to raise a huge interest. I do not want to discuss any possible link between the fringes created by the electric field and the aura. The aim of the paper is simply that of discussing the researches by Fernando Sanford on electric photography, documented by images on a scientific journal. To the author's knowledge, the picture of the left panel of Figure 1 is one of the early images of the Kirlian effect, that we can find in a journal accompanied by a clear discussion of the method used to have it. It was in 1894: after many years, the electric fringes are still considered as an effect of aura. Unlike Kirlian, Sanford did his best to reduce these fringes, improving the device but losing his "Sanford effect".

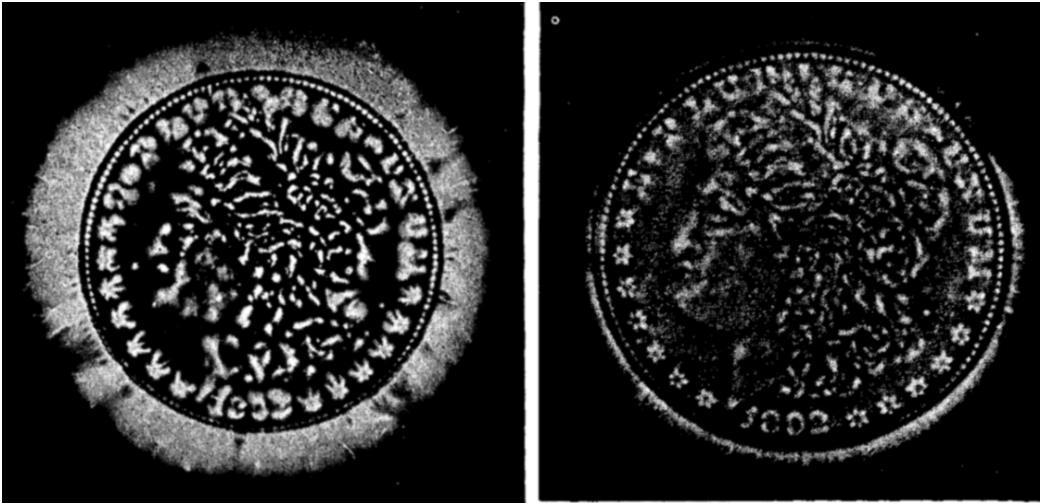

Fig.1. The images of a coin that Sanford obtained by means of his electric photography (adapted from the images in Ref.[3]). The left panel is showing the result of an earlier experiment and it is clearly possible to see the fringes. On the right, an improvement of the device reduced the fringes.

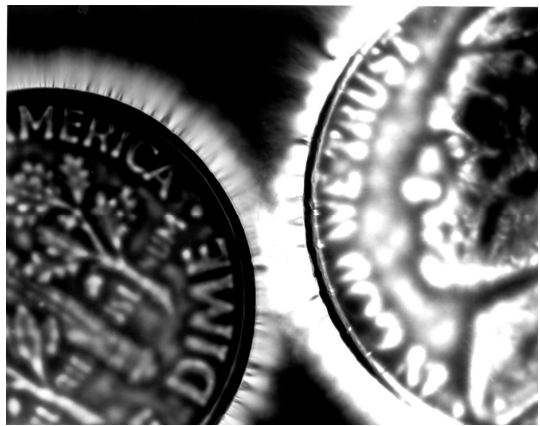

Fig.2. A Kirlian photography of coins, from Ref.6.